\documentstyle[12pt]{article}
\newcommand{\reali}{{\hbox{{\rm I}\kern-.2em\hbox{\rm R}}}}
\newcommand{\complessi}{{\ \hbox{{\rm I}\kern-.6em\hbox{\bf C}}}}
\newcommand{\rf}[1]{(\ref{#1})}
\newcommand\beq{\begin{equation}}
\newcommand\eeq{\end{equation}}
\newcommand\beqy{\begin{eqnarray}}
\newcommand\eeqy{\end{eqnarray}}
\newcommand\de{\partial}
\newcommand{\eg}{{\em e.g.}}
\newcommand{\ie}{{\em i.e.}}
\newcommand{\hol}{{\rm Hol}}
\newcommand{\tr}{{\rm Tr}}
\newcommand{\xnot}{{x_0}}
\newcommand{\ave}[1]{\left\langle{#1}\right\rangle}
\newcommand{\intddue}[4]{\int_{#2}^{#1}d{#3}\int_{#2}^{#3}d{#4}}
\newcommand{\intdtre}[4]{\int_0^{#1}d{#2}\int_0^{#2}d{#3}\int_0^{#3}d{#4}}
\newcommand{\intdquattro}[5]{\intdtre{#1}{#2}{#3}{#4}\int_0^{#4}d{#5}}

\begin{document}
\begin{titlepage}
\title{Three-dimensional $BF$ Theories and\\
the Alexander--Conway Invariant of Knots\footnote{Work supported by
grants from MURST and by INFN}}
\author{
Alberto S. Cattaneo\\
Dipartimento di Fisica, Universit\`a di Milano\\
and I.N.F.N., Sezione di Milano\\
via Celoria 16, 20133 {\sc Milano, Italy}\\
E-mail: {\em cattaneo@vaxmi.mi.infn.it}
\and
Paolo Cotta--Ramusino\\
Dipartimento di Matematica, Universit\`a di Milano\\
and I.N.F.N., Sezione di Milano\\
via Saldini 50, 20133 {\sc Milano, Italy}\\
E-mail: {\em cotta@vaxmi.mi.infn.it}
\and
Maurizio Martellini\\
Dipartimento di Fisica, Universit\`a di Milano\\
and I.N.F.N., Sezione di Pavia\\
via Celoria 16, 20133 {\sc Milano, Italy}\\
E-mail: {\em martellini@vaxmi.mi.infn.it}
}
\date{July 13, 1994\\IFUM 478/FT\\ hep-th/9407070 }
\maketitle
\begin{abstract}
We study 3-dimensional $BF$ theories and define observables related
to knots and links. The quantum expectation values of these observables
give the coefficients of the Alexander-Conway polynomial.
\end{abstract}
\thispagestyle{empty}
\end{titlepage}
\newpage
\section{Introduction}

After Witten \cite{Witt} clarified
the existing relation between
Chern-Simons topological field theory and Jones link polynomials,
a significant amount of research in this subject has been developed.

{}From \cite{Witt} we learned that link polynomials with two variables
(the so called HOMFLY polynomials) can be recovered in the framework of
Chern-Simons quantum field theory, by considering the expectation
values of link observables, given (classically) by the product of the
traces of the holonomies computed along the components of the given
link.

More precisely the situation is as follows: the expectation values
of link-observables in a Chern-Simons field theory with gauge group $SU(N)$,
are related to the HOMFLY polynomial $P(l,m)$, provided that we require the
normalization condition $P(l,m)(\emptyset)=1$, where
$\emptyset$ is the empty knot (and not the unknot) and provided
(more fundamentally) that the two variables $m$ and $l$ are assigned some
specific values depending on the integer $N$ and on the
(quantized) coupling constant of the theory.

The Alexander-Conway polynomial
$\Delta(z)$ is the specialization of
the HOMFLY polynomial $P(l,m)$, characterized by the condition $l=1$
and $m=z$. These conditions are incompatible with the
condition arising from Chern-Simons field theory. This explains
why we cannot recover the Alexander-Conway polynomial in the
framework of Chern-Simons field theory.

Here we propose to consider a topological field theory in 3-dimensions,
that differs, in some key features, from the Chern-Simons theory
and that is related, as we shall see in this paper, to
the Alexander-Conway polynomial.
This field theory is called $BF$, from the symbols used to denote
the fundamental fields of the theory:
$F$ (or $F_A$) stands for the curvature of a connection $A$, while
$B$ is an
extra field which behaves, under gauge transformations, as the difference
of two connections.

A completely similar field theory can be defined in four dimensions, by
assuming that the field $B$ is a 2-form (instead of a 1-form) which behaves
like the curvature of a connection. Like the Chern-Simons theory,
$BF$ theories have been considered
extensively in the literature (\cite{Bla}, \cite{MS}) both in 3 and in 4
dimensions.

What has not been attempted so far, (a part from \cite{CM})
was to relate these topological
field theories with knot-invariants. Here we mean ordinary knots for
3-dimensional $BF$ theories and 2-knots for 4-dimensional $BF$ theories.

One of the reasons why this attempt has not been made, may well have to do
with the problem of finding a reasonable link- (or knot-)observable.
Let us elaborate a little bit on this point.

In a topological
field theory of the $BF$ type, one has to compute
functional integrals, by integrating over two basic fields: the connection $A$
and the $B$ field. This is a crucial difference with the Chern-Simons case,
where the only field involved is the connection $A$. The link-observables
should therefore contain not only the connection $A$, but also the field $B$.
In a $BF$ theory expectation values of link-observables containing only the
field
$A$ (like the Wilson loop operator) or only the field $B$ appear to be
trivial.

In this paper we propose a special kind of link-observable, which contains
both the field $B$ and the field $A$ (through the holonomy). More
precisely the observable is defined first by integrating along a knot $K$
the differential 1-form given at $y\in K$ by
$\hol_{\xnot}^y B(y) \hol_y^{\xnot}$,
where $\xnot$ is a fixed point on $K$,
and then by taking the trace of the exponential of the
integral defined above (in a given representation).

In a similar way, we can associate to any link $L,$
the product of the traces of the
exponentials of the above integrals, computed over all
the components of $L$.

In this way the knot and link observables of the $BF$ theory,
share the same basic
properties (including gauge invariance) with the Wilson line operators
considered by Witten, but
they depend on two different fields and moreover are expressed as
traces of exponentials of
{\it integrals of differential forms} along the knots,
i.e. no further path-ordering is required
besides the path-ordering
encoded in the definition of
the holonomy.

In this paper we are going to prove that the $BF$ theory
gives directly the Alexander-Conway polynomial in a variable $z$
that is proportional to the coupling constant of the theory.

Differently from the Chern-Simons case, no resummation is needed in
order
to recover the knot-polynomials. Here the terms of the perturbative
series are exactly the coefficients of the Alexander-Conway polynomial.

This is perfectly consistent with the fact that Feynman integrals,
in a topological field theory with knots incorporated,
define Vassil'ev invariants of the given knot, and that the only polynomial
whose coefficients are
Vassil'ev invariants of finite type,
is the Alexander-Conway polynomial \cite{BN1}.

In this way, the $BF$ theory is suggesting a way of computing (at least
theoretically) the coefficients of the Alexander-Conway polynomial
 of a given knot,
as multiple integrals over copies of the
knot. The kernels of these integrals are similar in nature, to
the ones considered in defining linking numbers and the Arf invariant,
but involve convolutions of higher orders (for a comparison,
one may see the discussion on
``Chinese characters" in \cite{BN1}).

One of the relevant
advantages of $BF$ theories is that they can be defined in any
dimension. The 4-dimensional case has been discussed in a
preliminary way, in
\cite{CM}. Even in that case one can in principle define invariants of 2-knots,
but the situation is far more complicated than the one considered in this
paper.

As far as physical applications are concerned, this paper can be considered
as a simplified model for a more complicated 4-dimensional theory.
The obvious hope
is that by considering topological field theories (\ie\
diffeomorphism-invariant field theories), both in 3 and 4 dimensions,
we can make some progress
in understanding the subtle appearance of the metric in
quantum gravity as the result of a broken phase of a topological
field theory.

A different relation between quantum fields and the Alexander-Conway
polynomial has been considered in \cite{Kau-sal}.

\section{BF-Theory in 3D: Action, Feynman Rules and Observables}

We consider a compact Riemannian 3-manifold $M$
and a (trivial) principal $SU(N)$-bundle
over $M$ with connection $A$ and curvature $F$.
Our quantum field theory will include
a field $B$, classically given by a
$su(N)$-valued
1-form of the adjoint type, \ie\ a 1-form given by the difference
of two connections.

We define the $BF$-action as:
\beq
S_{BF} = -{1\over 2\pi} \int_{M} Tr(B\wedge F)
=-{1\over 2\pi} \int_{M}\ d^3x\ \epsilon^{\mu\nu\rho}
\tr [B_\mu(\de_\nu A_\rho + A_\nu A_\rho)].
\label{sbf}
\eeq

As is well known, under a
gauge-transformation $g(x)$ the above fields transform as follows:
\beq
\begin{array}{ll}
A(x)&\rightarrow g^{-1}(x)A(x)g(x) + g^{-1}(x)dg(x),\\
B(x)&\rightarrow g^{-1}(x)B(x)g(x)\\
F(x)&\rightarrow g^{-1}(x)F(x)g(x).\\
\end{array}
\label{trasf}
\eeq

It is then evident that the action \rf{sbf} is gauge invariant.
Moreover
the action \rf{sbf} is also
diffeomorphism invariant.
These invariance properties, combined with the fact that the Lagrangian
does not depend on a background metric, can be summarized by saying
that we are dealing with a {\it topological quantum field theory}
(\cite{Wittop}, \cite{Bla}).

Moreover the action \rf{sbf} is also invariant under the following
special $BF$-transformation:
\beq
\begin{array}{ll}
B(x)&\rightarrow B(x)+d_A\chi(x)
\end{array}
\label{Btrasf}
\eeq
where $\chi$ is a
$su(N)$-valued
0-form of the adjoint type (\ie\
a 0-form that behaves like an infinitesimal
gauge transformation) and $d_A$ denotes the covariant derivative.

The Gibbs measure of our theory will be given by
$\exp [(-1/k)S_{BF}]$,
where $k$ is the coupling constant.

Differently from the Chern-Simons case, the gauge invariance of our
action ensures that (the inverse of)
{\it the coupling constant should not necessarily
be equal to $\sqrt{-1}$ times an integer}.

Also differently from the Chern-Simons case, we can rescale one of the
fundamental fields of the theory with the coupling constant. In this way
we are able to shift the coupling constant from the action into
the observables\footnote{Remark that a scalar times a $B-$field is still a
$B$-field, while a scalar times a connection, is not a connection any more.}.

{\it From now on we incorporate the
coupling constant in the field $B$, by setting $k B_{new}= B_{old}$}

In the following we shall use the Lie-algebra notation $A=\sum_a A^aT^a$ and
$B=\sum_a B^aT^a$, where $T^a$ are the generators of $su(N)$ satisfying
the following relations:
\beq
\begin{array}{ll}
[T^a, T^b] &= f_{abc} T^c,\\
\tr(T^aT^b) &= -{1\over2}\delta^{ab}.
\end{array}
\label{generators}
\eeq

Here $f_{abc}$ are the structure constants for $su(N)$ and a summation
over upper and lower indices is understood.

As is well know, the covariant quantization of this theory requires
the introduction of a gauge fixing (\eg\ the Landau gauge).
The gauge fixing necessarily requires the introduction of a
background metric. However, as shown in \cite{Bla}
the quantum action, given by
the sum of the action \rf{sbf} plus the gauge-fixing and the
Faddeev--Popov terms  has an
energy-momentum tensor $T_{\mu\nu}$
which can be written as a pure BRST variation:
$T_{\mu\nu}=[Q,t_{\mu\nu}]$, where $Q$ is the BRST charge and the
explicit form of $t_{\mu\nu}$ is irrelevant for our purposes.
Thus the expectation values of a gauge-invariant observable $\cal O$,
with a vanishing commutator with $Q$, computed on physical states,
\ie\ states annihilated by $Q$, turns out to be diffeomorphism invariant.

Since we are mainly interested in the perturbative treatment of \rf{sbf}, we
shall choose $M=S^3$ as ambient space and work in a given chart.
Namely we will be working locally in $\reali^3$, where
we will be able to choose a flat metric
$g_{\mu\nu}=\delta_{\mu\nu}$.

The starting point for the perturbative quantization of \rf{sbf} are the
Feynman rules, which are given by the propagators:
\beq
\ave{A_\mu^a(x)B_\nu^b(y)}
=\delta^{ab}\epsilon_{\mu\nu\rho} {(x-y)^\rho\over|x-y|^3}
\label{prop}
\eeq
and the 3-vertex:
\beq
V=-{1\over8\pi}\int\ d^3x \epsilon^{\mu\nu\rho} f_{abc}
B_\mu^a \,A_\nu^bA_\rho^c.
\label{vert}
\eeq

Notice that the previous expressions for the propagators
and the vertex, imply the first important property of the
perturbative expansion:
\begin{enumerate}
\item[{\bf P.1}] The correlation functions $<A^sB^k>$ are non trivial only if
$k\geq s$
\end{enumerate}
In \cite{MS} it is shown that
perturbation theory, in the Landau gauge, is finite. This is due
to the existence of an underlying supersymmetry, which is in turn a consequence
of the invariance of the action under both transformations \rf{trasf}
and \rf{Btrasf}.
In particular the two sets of Faddeev-Popov ghosts arising from the
invariance of the action under transformations \rf{trasf} and \rf{Btrasf},
contribute to
the expectation values of a gauge-invariant observable exactly by
cancelling the graphs obtained by contracting the fields
$B$ and $A$ inside vertex insertions.

In other words in the perturbative
expansion of the BF-theory, one does {\it not need to consider
Wick contractions inside vertex insertions.}
This imply the second important property of our perturbative expansion:
\begin{enumerate}
\item[{\bf P.2}] The correlation functions $<A^sB^k>$ are non trivial only if
$s\geq k/2$.
\end{enumerate}

The partition function of the $BF$ theory (without knots)
has been shown to
be related to
the analytical Ray--Singer torsion \cite{Bla}.

When we incorporate knots, the partition function should then
be related to
the torsion of the {\it exterior of the knot}, which in turn
is related to the Alexander-Conway polynomial \cite{CM}.
This observation convinced us that we should look for a
direct verification, in perturbation theory, of the claim
that the partition
function with knots incorporated, gives the Alexander-Conway
polynomial.

Up to now nothing has been done in constructing
explicitly the correlation functions for $BF-$theories
and discuss their topological interpretation,
as it has been partly  done for the
Chern-Simons (CS) theory. In proving the above claim concerning the
Alexander-Conway polynomial, we are filling this gap.

The first problem we face is the construction of the analogue of the Wilson
line observable for the BF theories. Here we have two
fundamental fields $A$ and $B$, so one must consider
observables containing both of them.

This construction is realized in few steps.
As in the case for the Wilson line operator we start by considering a
fixed representation $R$ of $SU(N)$ and
the holonomy operator associated to a path $\gamma$ connecting two point
$x$ and $y$
\beq
\hol_x^y(A;\gamma)\equiv{\cal P}\exp\left[{
\int_{\gamma(x,y)} dz^\mu\ A_\mu(z)}\right],
\label{defhol}
\eeq
where $\cal P$ denotes path-ordering.
We then combine the matrices
$\hol$ and $B$
(in the given representation $R$) and
construct a matrix-valued 1-form:
\beq
G_\mu(y;\gamma,\gamma')\equiv
\hol_\xnot^y(A;\gamma) B_\mu(y) \hol_y^\xnot(A;\gamma'),
\label{defg}
\eeq
where $\xnot$ is an arbitrary fixed point in $S^3$ and $\gamma, \gamma'$
are two smooth curves connecting $x_0$ and $y$ and, respectively,
$y$ and $x_0$\rlap. Under a
gauge transform $g(x)$ we have the following transformation rules:
\[
\hol_x^y(A;\gamma)\rightarrow g^{-1}(x) \hol_x^y(A;\gamma) g(y),
\]
and
\[
G(y;\gamma,\gamma')\rightarrow g^{-1}(\xnot) G(y;\gamma,\gamma') g(\xnot).
\]

We can now integrate the field $G$ along a knot $K$. We generally will choose
the point $\xnot$ to lie on $K$; with this choice,
the quantity $\displaystyle\oint_K G$
{\it transforms
under a gauge transformation, exactly
as $\hol_{\xnot}(A,K)$ }(in the given representation
$R$ \footnote{
This implies that the $n$-point function constructed by the $G$-fields is a
gauge-singlet, \ie
\[
\ave{G(x_1)\dots G(x_n)} =  f(x_1,\dots,x_n)\, {\bf 1}
\]
for a suitable scalar function $f$. Here we omitted the dependency of
the field $G$ on the paths $\gamma$'s.}.)

We are now able to associate,
for any given representation
$R$ of $SU(N)$, (gauge-invariant) observables to
any knot $K$ in $S^3$. For instance we may
consider, for any positive integer $n$,
\[
\displaystyle{\tr_R \left[\oint_K G(y;\gamma, \gamma')\right]^n}.
\]
More generally we will be interested in the ``series" of the above observables,
namely in the observable $W_R(K;k)$ given by:
\beq
W_R(K;k) := \tr_R\exp\left[{k\oint_K  G(y;\gamma,\gamma')}\right],
\label{obs}
\eeq
where $k$ is the coupling constant\footnote{We should remember that
we rescaled the $B$-field.}. In principle the observable $W_R(K;k)$ depends
also on the choice of $\gamma$ and $\gamma'$, but, as we shall see
later on, this dependency will not really matter (as far as
$\gamma \cup\gamma'$ and $K$ are unlinked). Hence we will not include
the paths $\gamma$ and $\gamma'$ among the variables on which our
observables explicitly depend.

It is important to notice that the observable \rf{obs} is also invariant under
the special transformations \rf{Btrasf}. In fact consider
$G$ defined as above and replace
$B$ with $d_A\chi$ in the definition of $G$. We assume here that
$\chi$ (which is of the same nature of an infinitesimal
gauge transformation) vanishes at the fixed point $\xnot$.
So we have:
\[
\begin{array}{ll}
\displaystyle{\oint_K \hol_\xnot^y(A;\gamma) d_A\chi
\hol^\xnot_y(A;\gamma')}=
\displaystyle{\oint_K  d\bigl(\hol_\xnot^y(A;\gamma) \chi
\hol^\xnot_y(A;\gamma')\bigr)}
\\
\\
=\bigl[\hol(A,\gamma\cup\gamma'),\chi(\xnot)\bigr]=0.
\end{array}
\]

In other words, when we set $B_t\equiv B + t d_A\chi,$ then we have:
\[
\displaystyle{{d\over dt}\bigg |_{t=0} W_R(K;k,t)=0},
\]
where $W_R(K;k,t)$ has been obtained by replacing $B$ with $B_t$
in $W_R(K;k).$

In principle we can associate to {\it each point} $y\in K$
a different pair of distinct
paths $\gamma, \gamma'$
to be included in the observable \rf{obs}. A natural choice
consists instead
of defining a knot $K_f$  infinitesimally close to the given
knot $K$, but {\it never intersecting
$K$}, so that $\gamma \cup \gamma' = K_f$. This choice will automatically
eliminate the divergences produced by the propagator when it
is evaluated at coincident points.

The knot $K_f$ is called  a {\it framing for the knot} $K$. In local
coordinates
the equation for $K_f$ can be given as follows:
\[
x^\mu(t)=y^\mu(t)+\epsilon n^\mu(t),\quad (\epsilon>0, |n(t)|=1),
\]
where $y^\mu(t)$ is a parametrization of $K$ and $n^\mu(t)$ is a
vector field normal to $K$. As far as the notation is concerned,
we will write $G(K_f)$ to denote the dependency of the field $G$ on the
framing $K_f$.

As we shall see in
the next section, the expectation value of the observable
\rf{obs} is invariant under a
deformation of the framing  $K_f$, provided that the knot $K$ is not
intersected.
Hence
the only residual dependence on $K_f$ lies in the {\it linking number}
$\hbox{ln}(K_f;K)$ between the knot and its framing.
We will consistently use the ``standard framing",
namely we will consistently require $\hbox{ln}(K_f;K)=0.$

Our basic aim is
to compute the normalized expectation value in perturbation theory:
\beq
\ave{K}_R(k) := {\ave {W_R(K;k)}\over\ave{W_R(\bigcirc;k)} },
\label{ev}
\eeq
where $\bigcirc$ denotes the unknot, and the expectation
value is given by a
functional integration with respect to the
Gibbs measure given by $\exp (-S_{BF})$.

It is precisely \rf{ev} that gives the
Alexander-Conway polynomial, as we are going to show.

\section{Framing Invariance and Surgery: Preliminary
Non-Perturbative
Results}

As we already mentioned,
care should be exercised in dealing with the framing $K_f$.

In quantum
Chern-Simons theory one really finds invariants of framed
knots. Hence it is
perfectly natural to ask whether the same situation occurs here.
This is not the case. The basic idea is that in the $BF$-theory
we choose the standard framing
at the very beginning, and we stick to this choice
in all our calculations. The choice of the framing is part of the
definition of the observables
and tautologically {\it the framing-independence is
guaranteed at a non-perturbative level}.
In Chern-Simons theory, on the contrary, the
framing-dependent regularization
has to be assigned order by order in the perturbative
calculation (\cite{GMM},\cite{BN}).

In order to have a better understanding
of the framing-independence
of our quantum field theory,
we consider the effect on our observable
of a deformation
of $K$ and $K_f$ localized at a given point $x$, with coordinates
$x_{\mu}$
(see \cite{CGMM} for a related approach).

The following identities must be taken into
account:
\beq
\begin{array}{ll}
\displaystyle{{\delta\over\delta K^\mu(x)} \oint_K G} &=
\dot K^\nu(x)
\hol_\xnot^{x} (d_A\,B)_{\mu\nu}(x)
\hol_{x}^\xnot,\\
\displaystyle{{\delta\over\delta K_f^\mu(x)} \hol_w^z(K_f)} &=
\dot K_f^\nu(x) \hol_w^{x} F_{\mu\nu}(x) \hol^z_{x}.
\\
\end{array}
\label{var}
\eeq
$\dot K^\nu(x)$ and $\dot K_f^\nu(x)$ are the tangent
vectors to the knot $K$ and, respectively, to its framing $K_f$.

The functional derivatives of the $BF$
action are as follows:
\beq
\begin{array}{ll}
-\displaystyle{{\delta S\over\delta B_\mu^a(x)}} &=
\displaystyle{{1\over 8\pi}\epsilon^{\mu\nu\rho} F_{\nu\rho}^a(x)},\\
-\displaystyle{{\delta S\over\delta A_\mu^a(x)}} &=
\displaystyle{{1\over 8\pi}\epsilon^{\mu\nu\rho} (d_A\,B)_{\nu\rho}^a(x)}.
\\
\end{array}
\label{svar}
\eeq

In force of equations \rf{svar}, we can represent $F$ and $d_AB$, appearing
in the vacuum expectation value
of any observable,
as functional derivatives of the Gibbs measure $\exp (-S_{BF})$.
In particular
the variation with respect to $K$ ($K_f$) of the vacuum expectation value of
the observable $W$, can be replaced by a
functional derivative with respect to $A$ ($B$).

An integration by parts allows us to shift
this derivative to the remaining part of
the observable. The scheme is as follows:
\beq
\begin{array}{ll}
\displaystyle{{\delta\over \delta K(x)}} \longrightarrow d_AB(x)
\longrightarrow
&\displaystyle{{\delta\over\delta A(x)}},\\
\displaystyle{{\delta\over \delta K_f(x_f)}}
\longrightarrow F(x) \longrightarrow
&\displaystyle{{\delta\over\delta B(x)}},
\\
\end{array}
\label{parts}
\eeq

Thus a deformation of $K$ ($K_f$) gives a contribution only if the
functional derivative with respect to $A$ ($B$) is not zero. Since $A$ ($B$)
lives on $K_f$ ($K$), this is possible only if the deformation of $K$
path intersects $K_f$, namely only if
$\hbox{ln} (K;K_f)$ is changed. But we have to stick to the standard framing,
\ie\ no
modification of $\hbox{ln} (K;K_f)$
is allowed. We conclude that ${\ave{K}}_R(k)$
defines a true knot invariant.

As a preliminary non perturbative computation we now derive a
``surgery formula" for ${\ave K}_R(k)$. For this purpose we need to recall
some mathematical background concerning
the ``tangles'' in knot theory.

A tangle is obtained by a link (knot) diagram by breaking two edges
as
in fig.1.
One can  sum two tangles $\cal A$ and $\cal B$, as in fig.~2,
by forming the tangle
$\cal A+B$ in which the right outer strings of $\cal A$ and
the left outer strings of $\cal B$ are joined in agreement with their
orientation. One may recover a link diagram from a tangle in the two
ways described in fig.3.
The two link diagrams above are denoted respectively by the symbols
${\cal A}^N$ and ${\cal A}^D$ where the superscript $N$ and $D$
stands for ``numerator" and ``denominator". The terminology here is due to
Conway \cite{Con}.
It can be easily checked
that if $A^N$ ($A^D$) is a knot diagram, then $A^D$ ($A^N$)
is the diagram of a two-component link.

We know \cite{Lic} that when
$P$ is  the two variable HOMFLY polynomial or
any specialization of it (like the Alexander-Conway polynomial)
then for any tangles $\cal A$ and $\cal B$ we have:
\beq
P[({\cal A+B})^D] = P[{\cal A}^D] P[{\cal B}^D],
\label{den}
\eeq

Here $P$ is normalized so that
$P(\bigcirc)=1$.

We shall now show that the condition \rf{den} is actually satisfied by our
{\it normalized} knot invariant $\ave{K}_R(k)$, \ie\ the
knot invariant $\ave{W_R(K;k)}$, divided by
$\ave{W_R(\bigcirc;k)}$.

Namely we want to show that in quantum field theory the following relation
holds:
\beq
\ave{W_R(({\cal A+B})^D;k)}\ave{W_R(\bigcirc;k)} =
\ave{W_R({\cal A}^D;k)}\ave{W_R({\cal B}^D;k)}.
\label{denw}
\eeq

We assume that $A^D$ and $B^D$ are knot-diagrams, or equivalently that
$({\cal A+B})^D$ is a
knot-diagram.

The main ingredient in the proof is the cluster
property of the vacuum expectation values,
which allows to rewrite the l.h.s of equation
\rf{denw} as:
\[
\ave{W_R(({\cal A+B})^D;k)}\ave{W_R(\bigcirc;k)} =
\ave{W_R(({\cal A+B})^D;k) W_R(\bigcirc;k)}
\]
where $({\cal A+B})^D$ and the unknot  are supposed, for the purposes
of quantum field theory, to be at an infinite
distance. By using the diffeomorphism invariance of the $BF$
theory we may move the unknot $\bigcirc$ over $({\cal A+B})^D$  as shown in
fig.4.
Furthermore we are free to move the
companion of the unknot, denoted here by the symbol: $\bigcirc_f$,
independently of $\bigcirc$, as far
we keep $\bigcirc_f$ and
$\bigcirc$ unlinked.
Hence the l.h.s of equation \rf{denw} becomes:
\[
\begin{array}{ll}
\Biggl\langle
\tr\exp\left[
k\displaystyle{
\left(\int_{A'}+\int_{B'}\right) G(K_f=A'_f + B'_f)}\right]\\
\times \tr\exp\left[
k\displaystyle{
\left(\int_{U_1}+\int_{U_2}\right) G(\bigcirc_f=U_{1f} + U_{2f})}
\right]
\Biggr\rangle.\\
\end{array}
\]

Here $A'$ ($B'$) are the paths obtained by splicing together the strings of
$\cal A$ ($\cal B$) which are not spliced in the sum $\cal A+B$
and $U_{1,2}$ is a suitable
decomposition of $\bigcirc$ as shown in fig.5.
By using again the diffeomorphism
invariance, we take $A'$ and $B'$ infinitely apart, so that when $G$ is
integrated over $A'$ ($B'$) we may neglect the contribution coming from
$B'_f$ ($A'_f$) and replace it by $U_{1f}$ ($U_{2f}$). We can then repeat
this operation for $\bigcirc$ and the l.h.s. of equation \rf{denw} becomes:
\[
\begin{array}{ll}
\Biggl\langle
\tr\exp\left[
\displaystyle{k\int_{A'}G(A'_f+ U_{1f}) +
k\int_{B'}G(U_{2f}+ B'_f)}\right]\\
\times
\tr\exp\left[\displaystyle{
k\int_{U_1}G(A'_f+ U_{1f}) +
k\int_{U_2}G(U_{2f}+ B'_f)}\right]
\Biggr\rangle.\\
\end{array}
\]

Since averages of $G$-fields are in a gauge-singlet representation, it
follows that one is allowed to replace the traces with the dimensions of the
representation. Moreover, we can freely commute two G-fields
defined on two widely separated points. Both of these properties imply
that one can treat the arguments of the exponentials as
Abelian-like fields. Therefore the l.h.s. of equation \rf{denw} becomes:
\[
\begin{array}{ll}
\Biggl\langle
\tr\exp\left[
k\left(\displaystyle{\int_{A'}+\int_{U_1}}\right)
G(K_f=A'_f + U_{1f})\right]\\
\times \tr\exp\left[
k\left(\displaystyle{
\int_{U_2}+\int_{B'}}\right) G(K_f=U_{2f} + B'_f)\right]
\Biggr\rangle=\\
\ave{W_R({\cal A}^D;k)W_R({\cal B}^D;k)},\\
\end{array}
\]
where we have used the identities $A'+U_1= {\cal A}^D$ and $U_2 + B'={\cal
B}^D$.
By using again the diffeomorphism invariance
and the cluster properties the l.h.s. of equation \rf{denw} finally becomes:
\[
\ave{W_R({\cal A}^D;k)}\ave{W_R({\cal B}^D;k)
},
\]
namely we have proved equation \rf{denw}.
\section{Perturbative Expansion}

Let us consider the perturbative expansion of $\ave {W_R(K;k)}$
in powers of $k$.
At the $n-th$ order in $k$ we have a product of $n$ factors
$\displaystyle{\oint
G}$. Then, by taking into account
the structure of the $G$ operator in \rf{defg}, we have
to compute a correlation function with $n$ $B$-fields and an arbitrary
number of $A$-fields coming from the (path-ordered) expansions of the
holonomies. Furthermore we have to take into account
the properties {\bf P.1} and {\bf P.2} of the perturbative series,
that we derived in section {\bf 2}.

When fields are evaluated at coincident points, we could have,
in principle,
non-analytic correlation functions. But the structure of
$BF$ theories does not present this problem. In fact we have:
\begin{itemize}
\item one possible source of divergence  given by the propagator $\ave
{AB}$. This divergence does not appear since
$B$-fields live on the knot $K$ while
$A$-fields live on the framing $K_f$
and $K_f$ does not intersect $K$.
\item another possible source of divergence coming from a vertex
insertion. Again this divergence does not appear since
the structure of the observables implies that between two $B$-fields
there exists always an $A$-field which forbids them to
be evaluated at coincident points.
\end{itemize}

In order to calculate the terms of
the perturbative expansion, we
have to compute, according to the Wick
theorem,  the convolution of two and three point correlation
functions of the form. Since we do not have
contractions inside vertex insertions, these correlation functions
are given by:
\beq
\begin{array}{ll}
\ave{A_\mu^a(x)B_\nu^b(y)} &=
4\pi \, l_{\mu\nu}(x,y)\, \delta^{ab},\\
\ave{B_\rho^a(z)A_\nu^b(y)B_\mu^c(x)} &= (4\pi)^2\, v_{\mu\nu\rho}(x,y,z)
\, f^{abc},\\
\end{array}
\label{corr}
\eeq
where, in force of  \rf{prop} and \rf{vert},
$l$ and $v$ are explicitly given by
\beq
l_{\mu\nu}(x,y) = {1\over4\pi} \epsilon_{\mu\nu\rho}
{(x-y)^\rho\over|x-y|^3},
\label{defl}
\eeq
and
\beq
v_{\mu\nu\rho}(x,y,z) = \epsilon^{\alpha\beta\gamma}
\int_{S^3} d^3w\ l_{\mu\alpha}(x,w)\, l_{\nu\beta}(y,w)\,
l_{\rho\gamma}(z,w).
\label{defv}
\eeq
We will refer to \rf{defl} as to the {\it $l-$kernel} and to \rf{defv} as to
the {\it $v-$kernel}.
These kernels have the following obvious symmetries:
\beq
\begin{array}{ll}
l_{\mu\nu}(x,y) &= l_{\nu\mu}(y,x) ,\\
v_{\mu\nu\rho}(x,y,z) &= v_{\nu\rho\mu}(y,z,x) =
v_{\rho\mu\nu}(z,x,y),\\
v_{\mu\nu\rho}(x,y,z) &= -v_{\rho\nu\mu}(z,y,x).\\
\end{array}
\label{symm}
\eeq

Since the correlators \rf{corr} have to be integrated over the knot and
its framing, it is useful to define the following loop-dependent kernels
\beq
\begin{array}{ll}
l_K(s_1,s_2) &:= \dot K^\mu(s_1)\,\dot K^\nu_f(s_2) \
l_{\mu\nu}(K(s_1),K_f(s_2)),\\
v_K(s_1,s_2,s_3) &:= \dot K^\mu(s_1)\,\dot K^\nu_f(s_2) \, \dot K^\rho(s_3)
\ v_{\mu\nu\rho}(K(s_1),K_f(s_2),K(s_3)),\\
\end{array}
\label{deflv}
\eeq
where $K(\cdot)$ and $K_f(\cdot)$ $\colon [0,1]\to S^3$
are parametrizations of the knot and of its framing. As an
immediate consequence
of the property \rf{symm} we have:
\beq
v_K(s_1,s_2,s_3)=-v_K(s_3,s_2,s_1),
\label{loopsymm1}
\eeq
Moreover, in the limit when the spacing between $K$ and $K_f$ goes to zero
(without modifying the standard framing), we have also:
\beq
v_K(s_1,s_2,s_3)=v_K(s_2,s_3,s_1).
\label{loopsymm2}
\eeq
One of the main results of the present section is to prove the following
peculiar feature of our theory:
\begin{enumerate}
\item[{\bf P.3}] The terms in the
perturbative expansion, which contain an odd number of $B-$fields vanish.
\end{enumerate}

Indeed at these orders in perturbation theory, we have:
\begin{enumerate}
\item
amplitudes which directly allow a factorization of the
term $L(K,K_f):=
\displaystyle{\int_0^1 ds_1\int_0^1 ds_2\ l_K(s_1,s_2)}$. But this term
is the (Gauss formula for
the) linking number $\hbox{ln}(K_f;K)$, that is identically zero,
due to our choice of the standard framing.
\item
amplitudes corresponding to Feynman
graphs of the form depicted in fig.6.  These amplitudes are
computed by requiring
$K_f$ to be kept apart from $K$
at a distance  $\epsilon$. Then we send $\epsilon$ to zero.
As we will show at the end of this section,
this allows us to use the symmetries \rf{loopsymm2} and to show that also
these amplitudes are identically zero.
\end{enumerate}
Let us now analyze first the structure of terms of even order in
the perturbative expansion.

The terms of order $k^{2n}$ include a set of graphs given only by
convolution of $v-$kernels,
\ie\ containing exactly $n$ $A$-fields as in fig.7.
We
shall call these graphs ``V-graphs". Their structure is of the form
\beq
W^V :=\ave{(B^2A)^n}.
\label{Vgraph}
\eeq

At the $2n-th$ perturbative order, there
exist other Feynman graphs, of the type
\beq
\ave{B^{2n} A^{n+s}}, \qquad n\geq s>0
\label{othergraph}
\eeq
obtained by inserting (and Wick-contracting)
$s$ $A$-fields in the graphs \rf{Vgraph}.The insertion of one $A$-fields
implies the replacement
of a $v-$kernel in the graphs \rf{Vgraph}  with a pair of $l-$kernels.

Concerning the Lie
algebra factors for the graphs of
order $2n$, we have shown that, up
to the fourth order, it is always given by $(c_vc_2(R))^n$.
Here $R$ is the given representation of $SU(N)$, $c_2(R) {\bf 1}$
is its Casimir operator (\ie\ $\sum_a R(T^a)R(T^a)$)
and $c_v$ is defined by:
\beq
f^{acd}f^{bcd}=c_v\delta^{ab}.
\label{defcv}
\eeq

We conjecture that at any perturbative order $2n$,
all graphs have
$(c_vc_2(R))^n$ as a common factor. That means that the true
expansion parameter must be:
\beq
z^2 := (4\pi k)^2\, c_vc_2(R).
\label{defz}
\eeq

In order to justify the above conjecture,
we notice that if there existed Feynman amplitudes with different Lie algebra
factors,
then we would be able to
construct a multivariable knot-invariant. But this will
be incompatible with the skein relation (that we will prove in section {\bf
6}).

Hence (by taking into account property
{\bf P.3}), our perturbative
expansion will look like:
\beq
{\ave W_R(K;k)} = \dim(R) \sum_{n=0}^\infty w_{2n}(K) z^{2n},
\label{expw}
\eeq
for some suitable invariants $w_{2n}(K)$ of the knot $K$, {\it with
no residual dependence on the group itself} \footnote{Instead in the CS
theory the analogous of the $w_{2n}$ coefficients have an explicit
dependence on the group.}. Here $z$ is given as in \rf{defz}
and $\dim (R)$ is the dimension of the given representation.

We want now to
construct explicitly the coefficients $w_{2n}(K).$

Let us first consider graphs of the type \rf{Vgraph}, i.e. graphs
given by convolutions of $n$ $v-$kernels.
They can be of two types: connected and non-connected.
The connected ones $w^V_{2n}(K)$ are defined as;
\beq
\begin{array}{ll}
w_{2n}^V(K)\equiv
\displaystyle{\int_0^1 ds^1_1
\int_0^{s^1_1} ds^1_2\int_0^{s^1_2} ds^1_3
\int_0^1 ds^2_1 \int_0^{s^2_1} ds^2_2\int_0^{s^2_2} ds^2_3}
\cdots\\

\times
\displaystyle{\int_0^1 ds^n_1 \int_0^{s^n_1} ds^n_2 \int_0^{s^n_2} ds^n_3
\prod_{i=1}^n
v_K(s_1^i, s_2^{i-1}, s_3^i).}
\label{wertex1}
\end{array}
\eeq

Here we have set: $s_2^0\equiv s^n_2.$ In equation \rf{wertex1} we can
alternatively
replace:

\noindent
$\displaystyle{
\int_0^1 ds^i_1 \int_0^{s^i_1} ds^i_2 \int_0^{s^i_2} ds^i_3 }$
with
$\displaystyle{{1\over 2}
\int_0^1 ds^i_1 \int_0^1 ds^i_3 \int_{s^i_3}^{s^i_1} ds^i_2 .}$

In equation \rf{wertex1}, one can immediately notice
that each $v-$kernel is ``linked" to the next one, and so there is only
one ``chain" of ``linked" $v-$kernels, this is exactly the meaning of the word
``connected". Non-connected V-graphs are
V-graphs containing more than
one chain of ``linked" $v-$kernels.

Up to the fourth order in perturbation theory, the non-connected
V-graphs do not appear and it is reasonable to expect that
the same will be true
at any order of perturbation theory. We will therefore consider only connected
V-graphs, but it should be pointed out that this is
not a serious restriction.
In other words, all the arguments we developed and are going to develop,
will work equally well (with minor modifications) even if non trivial
non-connected V-graphs existed.

Now we are going to discuss a method that will enable us to
produce all the
terms of any given order of the perturbation series, from the
V-graphs of the same order.

In order to do this we will use the following ``rule" for Wick-contractions
(denoted by over/underlines):

\begin{equation}
\newcommand{\bac}{{\backslash}}
\newcommand{\hp}{\hphantom}
\newcommand{\ovl}{\overline}
\newcommand{\unl}{\underline}
\begin{array}{ccccc}
\ovl{\bac \hp{B\ldots/ A'A}/}\hp{\ldots B'\bac }
&\hp=&
\ovl{\bac \hp{B\ldots A}/}\hp{/ A'\ldots B'\bac }
&\hp+&
\ovl{\bac \hp{B\ldots[/ A',A}/}\hp{]\ldots B'\bac }\\
\hp\bac B\ldots\hp/ A'A\hp/\ldots B'\hp\bac
&=&
\hp\bac B\ldots A\hp/\hp/ A'\ldots B'\hp\bac
&+&
\hp\bac B\ldots[\hp/ A',A\hp/]\ldots B'\hp\bac \\
\hp{\bac B\ldots}\unl{/\hp{ A'A/\ldots B'}\bac }
&\hp=&
\hp{\bac B\ldots A/}\unl{/ \hp{A'\ldots B'}\bac }
&\hp+&
\hp{\bac B\ldots[}\unl{/ \hp{A',A/]\ldots B'}\bac }
\end{array}
\label{rule}
\end{equation}

When we sum the first term of the r.h.s. of \rf{rule} with the term
which is equal to the l.h.s. of \rf{rule},
but has two contractions interchanged, we obtain a contribution that
does not require a path-ordering of $A$ and $A'$. By iterating this procedure
we produce at the end linking numbers between $K$ and $K_f$ that are zero.
The rule \rf{rule} has a nice diagrammatical interpretation as shown in fig.8.

The perturbative series is then constructed out of terms like the second
part of the r.h.s. of \rf{rule}. Each one of these terms can be obtained by
a term which has one field $A$ less (but has the same
Lie algebra factor). Namely each one of these
terms is obtained by replacing a $v-$kernel with two $l-$kernels.
We will give then an analytic description of this procedure,
by introducing an operator $\cal D$ that
changes a
$v-$kernel into the products of two $l-$kernels.
Before doing so
we would like to notice, as a side remark, that fig.8 has a strong resemblance
with one of the rules considered in the computation
of Vassil'ev knot invariant (of finite type) \cite{BN1}.
This is not surprising, since at any fixed order of perturbation
theory, we expect that Feynman integrals of
topological field theories (with knots incorporated) will produce Vassil'ev
invariants\footnote{In CS theory the situation
was considerably different, since
a redefinition of the coupling constant $k\to t=\exp(2\pi\, i N/k)$ was
needed. This redefinition
implied an infinite  resummation of the Feynman graphs.}.

Let us now define the operator $\cal D$, that
we mentioned above.

The operator $\cal D$ is assumed to transform the
term $v_K(s_1^i, s_2^{i-1}, s_3^i)$ in \rf{wertex1},
according to the following rule:
\beq
\begin{array}{ll}
({\cal D}v_K)(s_1^i, s_2^{i-1},s_2^i, s_3^i)\equiv
&\displaystyle{{1\over 2}\int_{s^i_2}^{s^i_1} d\bar s_2
 \,l_K(s^i_1,s_2^{i-1})l_K(s^i_3,\bar s_2)}\cr
& +\displaystyle{{1\over 2}\int_{s^i_3}^{s^i_2}
d\bar s_2\,l_K(s_1^i,\bar s_2)\,l_K(s^i_3, s_2^{i-1})}
.\cr
\label{defD1}
\end{array}
\eeq

Notice that both
$v_K$ and ${\cal D}v_K$ change sign when we exchange
$s^i_1$ with $s^i_3$ .

So far we have defined the
action of the operator $\cal D$ on a single $v-$kernel.
But \rf{wertex1} is given, in general,
by the convolution of many $v-$kernel. The
action of the operator $\cal D$ on  \rf{wertex1} is then completely
defined
by assuming that $\cal D$ satisfies the {\it Leibniz rule}.

In a similar way we can
define the action of $\cal D$ on $w^V_{2n+1}(K)$,
i.e. on the connected graph of order
$2n+1$ given by the convolution of $n$ $v-$kernels
and one $l-$kernel.
This graph, which is of the type $\ave{B^{2n+1}A^{n+1}},$ is defined as:
\beq
\begin{array}{ll}
w^V_{2n+1}\equiv
\displaystyle{\int_0^1 ds^0_1 \int_0^1 ds_2^0\int_0^1 ds^1_1
\int_0^{s^1_1} ds^1_2\int_0^{s^1_2} ds^1_3
\int_0^1 ds^2_1 \int_0^{s^2_1} ds^2_2\int_0^{s^2_2} ds^2_3}
\cdots\\
\\
\times
\displaystyle{\int_0^1 ds^n_1 \int_0^{s^n_1} ds^n_2 \int_0^{s^n_2} ds^n_3
l_K(s_1^0,s_2^n)\prod_{i=1}^n
v_K(s_1^i, s_2^{i-1}, s_3^i). }
\label{wertex2}
\end{array}
\eeq

In a completely similar way we can define
the action of $\cal D$ on any kernel given by the convolution
of an arbitrary number of $v-$kernels and $l-$kernels.
The basic idea is always that any given $v-$kernel $v_k(x,y',z)$ is transformed
into:
\beq
({\cal D}v_K)
(x,y,y',z) = \frac12 \int_z^y d\bar y\,l_K(x,\bar y)l_K(z,y')+
\frac12 \int_y^x d\bar y\,l_K(x,y')l_K(z,\bar y),
\label{defD}
\eeq
where the variable $y$ is the same variable
appearing in the other $v-$ or $l-$kernel to which
the given $v-$kernel is ``linked".
A graphical description of the action of the operator $\cal D$
is given in fig.9.

Together with the operator $\cal D$, we can consider its exponential,
\ie\
the operator
$\exp{\cal D}$, defined, in the standard way, as a power series.
Notice that by applying
$\cal D$ a number of times greater than the number of
$v-$kernels appearing in a given graph, one obtains zero
\footnote{This fact is directly connected to property
{\bf P.1} of $BF$ theory.}. So the operator
$\exp{\cal D}$ is given, for any assigned graph, only by
a finite sum of powers of
$\cal D$.

The perturbative expansion of our $BF$-theory satisfies the following rules:
\begin{enumerate}
\item all the terms of order $2n$
of the perturbative expansion are obtained
by applying the operator
$\displaystyle{e^{\cal D}}$ to the graph \rf{wertex1}\footnote{
And
to its
non-connected counterparts,  if they exist.}.
\item all the terms of order $2n+1$
of the perturbative expansion are obtained
by applying the operator
$\displaystyle{e^{\cal D}}$ to the graph \rf{wertex2}.
\end{enumerate}

In order to prove the property
{\bf P.3}, it is sufficient to show that
\rf{wertex2} is zero, thus implying that any other graph obtained from
\rf{wertex2} by applying the operator $\cal D$ is also zero
\footnote{This proof applies equally well to any non-connected
counterpart of \rf{wertex2}}.

Due to the special symmetry \rf{loopsymm2}, obtained
in the limit when the spacing between the knot and its framing is sent
to zero, we have in \rf{wertex2}:
\[
v_k(s^1_1,s^0_2,s^1_3)=v_k(s^0_2,s^1_3,s^1_1)=v_k(s^1_3,s^1_1,s^0_2)
\]
and hence:
\beq
\begin{array}{ll}
3w^V_{2n+1}=
\displaystyle{{1\over 2}\int_0^1 ds^0_1 \int_0^1 ds_2^0\int_0^1 ds^1_1
\int_0^1 ds^1_3\left[\left(\int_{s^1_3}^{s_1^1} + \int_{s_1^1}^{s^0_2}
+\int_{s_2^0}^{s_3^1}\right)ds^1_2\right]}\\
\\
\times
\displaystyle{\int_0^1 ds^2_1 \int_0^{s^2_1} ds^2_2\int_0^{s^2_2} ds^2_3
\cdots
\int_0^1 ds^n_1 \int_0^{s^n_1} ds^n_2 \int_0^{s^n_2} ds^n_3 }\\
\times \displaystyle{l_K(s_1^0,s_2^n)\prod_{i=1}^n
v_K(s_1^i, s_2^{i-1}, s_3^i)=0. }
\label{wertex3}
\end{array}
\eeq

\section{Explicit Computations up to the fourth order}

By taking into account
\rf{expw}, the normalized expectation value \rf{ev}, will be
given by the following formula \beq
{\ave K}_R(k) = \sum_{n=0}^\infty a_{2n}(K)\, z^{2n},
\label{expK}
\eeq
where $z$ is defined in \rf{defz} and $a_{2n}(K)$ are to be determined as
a function of the coefficients $w_{2n}(K)$ and the corresponding
coefficients for the unknot $w_{2n}(\bigcirc)$.

More precisely we have;
\beq
\begin{array}{ll}
a_0(K) &= w_0(K),\\
a_2(K) &= w_2(K) - w_2(\bigcirc),\\
a_4(K) &= w_4(K) - w_2(\bigcirc)\,a_2(K) - w_4(\bigcirc),\\
\ldots&\\
a_{2n}(K) &= w_{2n}(K) - \sum_{i=1}^n w_{2i}(\bigcirc) a_{2n-2i}(K).\\
\end{array}
\label{defa}
\eeq

At order zero we have simply $w_0(K) = 1$.
At the second order we have only two graphs as shown in fig.10.
The V-graph (\ie\ $\ave {B^2A}$)  is given by
\beq
w_2^V(K) =
\intdtre1{s_1}{s_2}{s_3}\ v_K(s_1,s_2,s_3),
\label{wtwo}
\eeq
while the graph $\ave{B^2A^2}$ is given by:
\beq
{\cal D}w_2^V(K) =
\intdquattro1{s_1}{\bar s_2}{s_2}{s_3}
\ l_K(s_1,s_2)\, l_K(s_3,\bar s_2).
\label{dwtwo}
\eeq

The total second order term is then $w_2(K) = (1+{\cal D}) w_2^V(K)$.
As a consequence, $w_2(K)$ is the same as the second order
contribution of the perturbative expansion of
the CS theory, so that we can use the results of \cite{GMM}
and obtain;
\begin{enumerate}
\item $w_2(\bigcirc)= -{1\over 24}$;
\item $a_2(k) = w_2(K) + {1\over24}$
\end{enumerate}
As was pointed out in \cite{GMM}, the term $a_2(K)$ is
the so called Arf invariant, that is well known
to be the second coefficient of the Alexander-Conway
polynomial (\cite{Kauff}).
The third order of the perturbative expansion is trivial, thanks to
the results of the previous section. The structure of the
corresponding graphs is shown in fig.11.
We now consider the fourth order of the perturbative expansion.
The coefficient $a_4(K)$, is obtained by considering the Feynman diagrams
shown in fig.12.
An explicit computation shows that there exists only one V-graph ($\ave
{B^4A^2}$), which is completely connected (see fig.12a), namely:
\beq
w_4^V(K) =\frac14
\int_0^1ds_1 \int_0^1ds_3 \int_0^1dt_1 \int_0^1dt_3\,
\int_{s_3}^{s_1}ds_2 \int_{t_3}^{t_1}dt_2\
v_K(s_1,t_2,s_3)\, v_K(t_1,s_2,t_3).
\label{wfour}
\eeq

The contribution corresponding to fig.12b is given by:
\beq
\begin{array}{l}
{\cal D}w_4^V(K)
= \displaystyle{\frac14
\int_0^1ds_1 \int_0^1ds_3 \int_0^1dt_1 \int_0^1dt_3\,
\intddue{s_1}{s_3}{\bar s_2}{s_2}\, \int_{t_3}^{t_1}dt_2}\\
\\
\times \left[
l_K(s_1,t_2) l_K(s_3,\bar s_2) v_K(t_1,s_2,t_3)
+l_K(s_1,s_2) l_K(s_3,t_2) v_K(t_1,\bar s_2,t_3)\right],\\
\end{array}
\label{dwfour}
\eeq
while the contribution of fig.12c is given by
\beq
\begin{array}{l}
{1\over2}{\cal D}^2w_4^V(K)
= \displaystyle{\frac18
\int_0^1ds_1 \int_0^1ds_3 \int_0^1dt_1 \int_0^1dt_3\,
\intddue{s_1}{s_3}{\bar s_2}{s_2}\, \intddue{t_1}{t_3}{\bar t_2}{t_2}}\\
\\
\times \left[
l_K(s_1,t_2) l_K(s_3,\bar s_2) l_K(t_1,s_2) l_K(t_3, \bar t_2)
+l_K(s_1,\bar t_2) l_K(s_3,\bar s_2) l_K(t_1,t_2) l_K(t_3.s_2)
\right].\\
\end{array}
\label{ddwfour}
\eeq

The sum $(1 + {\cal D} + 1/2 {\cal D}^2)\, w_4^V(K)$
gives the term $w_4(K)$. In order to obtain the normalized fourth
order knot-invariant, one has to compute
$w_4(\bigcirc)$ for the unknot  and apply the relations \rf{defa}.

In principle \rf{defa} allows the computation of the
coefficients of the Alexander-Conway polynomial at any
order. We have in fact a
close analytic expression for such
coefficients. Actual computations may be difficult,
but the situation is considerably
simpler than the one in Chern-Simons theory. In CS theory, all the
terms of the perturbative series were framing-dependent, and a close
analytic expression for such terms
was not available.

\section{Skein Relation}

In the previous sections we proved
that the expectation values $\ave{K}_R(k)$ of our
knot-observables:
\begin{enumerate}
\item[a)] satisfy the (denominator) surgery formula
which is common to all known link polynomials (including the Alexander-Conway
polynomial)
\item[b)] are a power series (or a polynomial) in the variable:
$z := (4\pi k)\, \sqrt{c_vc_2(R)}$. The coefficients of these power series
are knot-invariants
\item[c)] coincide, up to second order,
with the Alexander-Conway Polynomial $\Delta_K(z)$
\item[d)] contain only terms
of even order in $z$
\end{enumerate}
We would like now to prove, in the framework of perturbation
theory, our claim that $\ave{K}_R(k)$ coincides at all orders with
the Alexander-Conway Polynomial.

Let us recall briefly the axiomatic definition of such polynomial.

For any link $L$, $\Delta_L(z)$ is a polynomial of one variable $z$
normalized so that $\Delta(\bigcirc)=1$. It is a link-invariant
satisfying the following skein relation:
\beq
\Delta_{K_+}(z) -\Delta_{K_-}(z) = z\Delta_{K_0}(z),
\label{skein}
\eeq
where $K_+,K_-,K_0$ are three oriented knots/links that are exactly the
same except near a crossing point where they look like as in fig.13.
The existence of such polynomial follows directly from the definition
of the classical Alexander-Conway invariant \cite{Con}. We recall here some
basic
properties of the Alexander-Conway polynomial:
\begin{enumerate}
\item the polynomial satisfying the normalization condition and the
skein relation defined above, is necessarily unique. Moreover
any knot-invariant given by a
(formal) power series satisfying the above skein relation,
{\it and} the normalization condition,
must necessarily be a polynomial
\item the above skein relation cannot be defined in terms of knots only.
In fact, for any link $L_+$
with $s$ components, $L_-$ has also $s$ components, but
$L_0$ has either $s+ 1$ or $s-1$ components
\item if a link $L$ is composed by two links separated
by a sphere, then the polynomial
$\Delta_L$ is zero,
\item if $H$ denotes the Hopf link (\ie\ the link given
by two linked circles, with linking number $+1$), then
$\Delta_H(z)=z$
\item for
any link $L$, equation \rf{skein} can be equivalently rewritten as:
\beq
a_{n+1}(L_+) - a_{n+1}(L_-) = a_n(L_0),
\label{skeina}
\eeq
where we have set $\Delta_L(z)\equiv \sum_{n\geq 0} a_n(L)z^n.$
\end{enumerate}
Further properties of the Alexander-Conway polynomial are
\begin{enumerate}
\item[6.]
for a knot $K$, $\Delta_K(z)$  is an {\it even} polynomial in $z$
\cite{Kauff}
\item[7.] for a 2-component link $L$,
$\Delta_L(z)$ is an {\it odd} polynomial in $z$
\item[8.]
\[
\begin{array}{ll}
a_0(K) &=\left\{\matrix{
1&\quad&\hbox{if $K$ is a knot}\hfill\\
0&\quad&\hbox{otherwise}\hfill\\}\right.\\
\\
a_1(L) &=\left\{\matrix{
\hbox{ln}(K^1;K^2)&\quad &\hbox{if $L=K^1\cup K^2$ is a two-component link}
\hfill\\
0&\quad&\hbox{otherwise}\hfill\\}\right.\\
\end{array}
\]
\end{enumerate}

As far as the coefficient $a_2(K)$ (Arf-invariant)
is concerned, we notice that we have
$a_2(K_+)-a_2(K_-)={\rm ln}(K^1_0;K^2_0)$,
where $K_0^1$ and $K_0^2$ are the two components of $K_0$. We are going to
recover directly this relation, in the framework of perturbation theory.

By property 2 above,
we know that, if we want to
recover a skein relation in the framework of perturbation
theory, we have to define $<W_R(L;k)>$ for a link $L$ which is not
necessarily a knot.

For any link $L$ with components $\{K_i\}_{i=1,2,\cdots,s}$
we first consider:
\beq
W_R(L;k)\equiv
\prod_i\tr\exp\left[
k\int_{K_i}G\right].
\label{conn}
\eeq

Instead of the expectation values $<W_R(L;k)>$ we will focus
here
on the {\it connected correlation functions} $<W_R(L;k)>_c$ .

The connected correlation functions can be
defined inductively as follows (see \eg\ \cite{ZJ}).

Let
$L$ be a link with $s$ components and let
${\cal P}(s) $ the set of all the
non-trivial partitions of the set $\{1,2,\cdots,s\}$.

For any such partition $\sigma\in {\cal P}(s),$ represented
by a multi-index $[\sigma_1,\sigma_2,
\cdots, \sigma_{k_{\sigma}}],$
with
\[
\sigma_1\cup\sigma_2\cup\cdots\cup\sigma_{k_{\sigma}}=
\{1,2,\cdots,s\},
\]
we consider the corresponding collection of links $L_{\sigma_1}$,
$L_{\sigma_2}$,$\cdots$, $L_{\sigma_{k_{\sigma}}}$.

Then we set, for a knot $K$:
\[
<W_R(K;k)>_c\equiv <W_R(K;k)>,
\]
and for a link $L$, with $s$ components:
\[
<W_R(L;k)>_c\equiv <W_R(L;k)>-\sum_{\sigma\in {\cal P}(s)}
<W_R(L_{\sigma_1};k)>_c
\cdots <W_R(L_{\sigma_{k_{\sigma}}};k)>_c
\]

The above definition implies:
\begin{itemize}
\item for any link $L=L_1\cup L_2$
composed by two {\it separate} links $L_1$ and $L_2$,
we have $<W_R(L;k)>_c=0.$ This is a direct
consequence of the cluster property and is consistent with property
3 above.

\item the (denominator) surgery law is satisfied. If $\cal A$ and $\cal B$
are two tangles,
then we have:
\[
\ave{W_R(({\cal A+B})^D;k)}\ave{W_R(\bigcirc;k)} =
\ave{W_R({\cal A}^D;k)}\ave{W_R({\cal B}^D;k)}.
\]
The proof of the above identity is {\it verbatim} the same we have considered
for the surgery of two knots. It can be easily seen that the same is true
also for the connected correlation functions.
\end{itemize}

We are now in position to recover the
skein relation \rf{skein}
for our link-observables.

We start by considering one knot with a selected crossing $K=K_-$ and
its switched counterpart $K_+$.

The switching can be seen as the result of applying a singular deformation
operator, namely:
\beq
\ave{W_R(K_+;k)} = \ave{W_R(K_-;k)}+\displaystyle
{{{\delta}\over {\delta v}}\ave{W_R(K_-;k)}},
\eeq
where $v$ is a vector (singular vector field)
based at the given crossing point of the link.
The singular deformation operator\footnote{This singular
deformation operator is a close relative of the derivation
introduced by Sossinsky \cite{Sos} for the computation
of Vassil'ev invariants.}  should in fact be applied to
{\it both the knot $K$ and the framing $K_f$}. So we can set:
\beq
\displaystyle{
{\delta\over {\delta  v}}}\equiv
{v}_\mu\left(\displaystyle{
{\delta\over {\delta K^{\mu}}}+
{\delta\over {\delta K_f^{\mu}}}
}\right).
\label{vartot}
\eeq

In switching  the crossing of both the knot and of the framing from
a crossing of type $-$ to a crossing of type $+$, we are changing the framing
of the original knot. In order to restore the standard framing, we need
to twist the pair $(K,K_f)$ as shown in fig.14.

We now use the integration by part techniques.
{}From now on, we consider the fundamental
representation of $SU(N)$. The generators $T^a$
satisfy \rf{generators} and also the:
Fierz identity:
\beq
\sum_a T^a_{ij}T^a_{kl} = {1\over2}
\left( \delta_{il}\delta_{jk} - {1\over N} \delta_{ij}\delta_{kl}
\right).
\label{fierz}
\eeq
When $R$ is the fundamental representation, then $c_2(R)$ will be
more simply
denoted by $c_2$.
The form of the $BF-$action
allows the following substitutions:
\beq
\begin{array}{ll}
\displaystyle{{\delta\over \delta K^{\mu}(x)}}
\longrightarrow
\displaystyle{{\delta\over\delta A_{\mu}(x)}},\\
\displaystyle{{\delta\over \delta K_f^{\mu}(x)}}
\longrightarrow
\displaystyle{{\delta\over\delta B_{\mu}(x)}},
\\
\end{array}
\eeq
where $x$ is the point where the (singular) deformation is performed.

Next we apply the above operators (derivatives) to our fields.
When we consider the variation of the holonomy, with respect to the
connection, we obtain the same formula obtained in the Chern-Simons
theory (\cite{CGMM}):
\beq
\displaystyle{
{\delta\over {\delta A^a_\rho(x)} }
}\hol_{\xnot}^{\xnot}
=
\displaystyle{
{\dot K_f^\rho(x)\, \hol_{\xnot}^{x} T^a\hol^{\xnot}_x}.
}
\label{varhol}
\eeq

When we consider instead the field $G$ that defines our $BF$ observables,
we obtain:
\beq
\begin{array}{ll}
\displaystyle{{\delta\over{\delta A^a_\rho(x)}} \oint G}
=&
\displaystyle{\int_{\xnot}^x \hol_{\xnot}^y B_{\sigma}(y)\hol_y^x
\dot K_f^\rho(x)\, T^a\,\hol_{x}^{\xnot}dy^{\sigma}}\cr\\
&+\displaystyle{\int_x^{\xnot}
\hol_{\xnot}^x \dot K_f^\rho(x)\,T^a\, \hol_x^y B_{\sigma}(y) \hol_y^{\xnot}
dy^{\sigma}}\cr\\
\\
\displaystyle{{\delta\over{\delta B^a_\rho(x)}} \oint G}
=&\displaystyle
\displaystyle{\hol_{\xnot}^x \dot K^\rho(x)\, T^a\hol_{x}^{\xnot}}.\cr\\
\label{varg}
\end{array}
\eeq

The action of
performing first the singular deformation of the knots and of its framing
{\it and then} integrating by parts (see \cite{CGMM}), can be
equivalently described
by the action of the following operators
on the vacuum expectation values:
\beq
\begin{array}{ll}
\displaystyle{{\delta\over \delta K^{\mu}(x)}}
&\rightarrow

\displaystyle{4\pi \epsilon^{\mu\nu\rho}\sum_a\left ({\delta^2\over
{\delta B^a_{\nu}(x(s))\delta A^a_\rho(x(\tilde s))}} -
{\delta^2\over{\delta B^a_{\nu}(x(s))\delta A^a_{\rho}(x(s))}}\right )}\\
\displaystyle{{\delta\over \delta K_f^{\mu}(x)}}
 &\rightarrow
\displaystyle{4\pi  \epsilon^{\mu\nu\rho}\sum_a\left ({\delta^2\over{
\delta A^a_{\nu}(x(s))\delta B^a_\rho(x(\tilde s))}} +
{\delta\over{\delta A^a_\nu(x(s))\delta B^a_\rho(x(s))}}\right )}\\
\end{array}
\label{exchab}
\eeq

Here $x$ is the crossing point of the knot,
while $x(s)$ and $x(\tilde s)$ denote the two distinct coordinates of the
knot for which we have: $x(s)=x(\tilde s)=x$.
The second terms in both the operators \rf{exchab} correspond to the twist
needed in order to restore the standard framing.

It is now apparent that the variation \rf{exchab} {\it reduces by one
the number of $B-$fields}. Namely when we apply the variation
\rf{exchab} to any term $w_{2n}(K),$ we obtain a term of order
\footnote{We recall that the number of $B-$fields
gives exactly the order of the given term
in the perturbative expansion.} $2n-1$ .

Moreover by applying the operators \rf{exchab} to any term $w_{2n}(K)$,
we create one matrix $T^a$ at the point $x(s)$ and one matrix $T^a$
at the point $x(\tilde s)$. This will happen both on the knot and on the
framing. Exactly as in \cite{CGMM}, the Fierz identity generates two
contributions. These are:
\begin{enumerate}
\item[$i$)] one contribution of order $2n-1$ relevant to the original
knot $K=K_-$. {\it This contribution is zero, due to the results of
section} {\bf 4}
\item[$ii$)]one contribution of order $2n-1$  relevant to the
2-component link $K_0$.
\end{enumerate}

Let us analyze in detail the characteristics of the second contribution $ii)$.
Whenever the two
components of the link $K_0$ are separated by a sphere,
then the term $ii)$ must be zero at any order.

Moreover since the vacuum expectation value of both $K_+$ and $K_-$
satisfy separately the denominator surgery formula (section {\bf 3}), then
the series obtained by summing all the contributions $ii)$
corresponding to the different orders in perturbation theory, must satisfy
the same denominator surgery formula.
This implies that the series obtained by summing all the
contributions $ii)$ must be proportional by a factor $f_2(z)$
to the series
obtained by summing the connected correlation functions.

There is an arbitrary choice to be made, concerning the term $f_2(z)$
considered
above. Any such choice is equivalent to a
choice of the normalization factor for
the expectation value (connected correlation functions)
of 2-component links.

In order to be consistent with the skein relation \rf{skein} (or
\rf{skeina}), we choose:
\beq
f_2(z)=\displaystyle{\frac {z \ave{W(\bigcirc;z/(4\pi \sqrt{c_v
c_2}))}}{\ave{W(H;z/(4\pi
 \sqrt{c_v c_2}))}_c} }
\eeq
This choice is equivalent to requiring that the $BF$ expectation value for
the Hopf link is exactly given by $z$.

{\it Thus we have proven, in the framework of perturbation theory,
the Alexander-Conway skein relation for the $BF$ theory.}

For any given integer $s$,
we could, in principle, choose a different normalization factor
$f_s(z)$ relevant to the (connected) vacuum expectation values of links
with $s$ components. If we want to preserve the denominator surgery
formula, then these normalization factors should satisfy
the following equation:
\[
f_s(z)=[f_2(z)]^{s-1},
\]
\ie\ only the factors $f_1(z)$ and $f_2(z)$ can be assigned independently.

As a conclusion we have the following relation between the Alexander-Conway
polynomial for a {\it link $L$ with $s$ components} and the (connected) vacuum
expectation values of the $BF$ theory:
\beq
\Delta_L(z)= f_s(z) \displaystyle{\frac{\ave{W_R(L;
z/(4\pi \sqrt{c_v c_2}))}_c}{\ave{W(\bigcirc;
z/(4\pi \sqrt{c_v c_2}))}}}.
\label{Alexander}
\eeq

As an example of the previous construction
we now perform the explicit computation of $\delta a_2(K)
\equiv a_2(K_+)-a_2(K_-)$.

We shall omit all the terms which
give the linking number of $K$ with $K_f$, since this has been
assumed to be zero. In particular the second halves of the two operators
\rf{exchab} can be omitted.

By taking into account \rf{exchab},\rf{varhol} and \rf{varg}, we obtain
\footnote{In order to avoid a cumbersome notation we write
$x(\tilde s)=\tilde x$ and $x(s)=x$ (see \rf{exchab})}:

\beq
\displaystyle{{\delta a_2(K)\over \delta K^{\mu}(x)}}=
{-\epsilon_{\mu\nu\rho}\dot K^\nu(x)
\over4\pi\,c_vc_2\,N }
\ave{ {\delta\over\delta A^a_\rho(\tilde x)}\,
\tr\hol_{\xnot}^x T^a\hol_x^{\xnot}\oint G},
\label{deltawd}
\eeq
and
\beq
\begin{array}{ll}
\displaystyle{{\delta a_2(K)\over \delta K_f^{\mu}(x)}}
&=
\displaystyle{{-\epsilon_{\mu\nu\rho}\dot K^\nu_f(x)
\over4\pi\,c_vc_2\,N}
\Biggl\langle {\delta\over\delta B^a_\rho(\tilde x)}\,}\\
&\displaystyle{
\tr\Biggl[\Biggl(
\int^x_{\xnot} dy^\sigma\ \hol^y_{\xnot} B_\sigma(y)\hol_y^x T^a\hol_x^{\xnot}
+}\\
&\displaystyle{+\int_x^{\xnot} dy^\sigma\ \hol^x_{\xnot} T^a \hol^y_x
B_\sigma(y)\hol_y^{\xnot}
\Biggr) \oint G\Biggr]\Biggr\rangle,}\\
\end{array}
\label{deltawdf}
\eeq
Then, by taking into account \rf{varhol}, \rf{varg}
and \rf{fierz}, we can rewrite
\rf{deltawd} and  \rf{deltawdf}:
as:
\newcommand{\ha}{\tr \hol^{\tilde x}_x}
\newcommand{\hb}{\tr \hol_{\xnot}^x \hol_{\tilde x}^{\xnot}}
\newcommand{\hc}{\tr \hol^{\xnot}_x\hol_{\xnot}^{\tilde x}}
\beq
\begin{array}{ll}
\displaystyle{v^{\mu}{\delta a_2(K)\over \delta K^{\mu}(x)}} &=
\displaystyle{{-1\over8\pi\, c_vc_2\,N}\Biggl(
\ave{\ha \, \tr \hol^x_{\xnot}\hol_{\tilde x)}^{\xnot}\oint G} +}\\
&\displaystyle{ +\ave{\hb\,\tr \int^{\tilde x} dy^\sigma\
\hol^y_\xnot B_\sigma(y)\hol_y^{\tilde x}\hol_x^{\xnot}}+}\\
&\displaystyle{
\ave{\hc\,\tr \int_{\tilde x} dy^\sigma\
\hol^x_\xnot \hol^y_{\tilde x}B_\sigma(y)\hol_y^\xnot}
\Biggr)},\\
\end{array}
\label{eqI}
\eeq
and
\beq
\begin{array}{ll}
\displaystyle{v^{\mu}{\delta a_2(K)\over \delta K_f^{\mu}(x)}} &=
\displaystyle{{-1\over8\pi\, c_vc_2\,N}\Biggl(
\ave{\ha\,\tr \hol^x_{\xnot}\hol_{\tilde x}^\xnot\oint G} +}\\
&\displaystyle{ +\ave{\hc\,\tr \int^{x} dy^\sigma\
\hol^y_\xnot B_\sigma(y)\hol_y^{x}\hol_{\tilde x}^\xnot}+}\\
&\displaystyle{
\ave{\hb\, \tr \int_{x} dy^\sigma\
\hol_\xnot^{\tilde x}\hol^y_{x}B_\sigma(y)\hol_y^\xnot}
\Biggr).}\\
\end{array}
\label{eqJ}
\eeq
Here the products $\epsilon_{\mu\nu\rho} v^\mu\dot K^\nu(x)\dot
K^\rho_f(\tilde x)$ and $\epsilon_{\mu\nu\rho} v^\mu\dot K^\nu_f(x)\dot
K^\rho(\tilde x)$ have both been normalized to one (as in Ref.
\cite{CGMM} and \cite{Bru}).
We now expand all the holonomies retaining only the linear part
in $A$. By taking into account
equation \rf{prop}, we have then to compute terms of the form:
\beq
\oint_\gamma dx^\sigma \oint_{\gamma'} dy^\rho\
\ave{\tr A_\sigma(x) B_\rho(y)}
= 4\pi \, \hbox{ln}(\gamma,\gamma')\, c_2 \,N,
\label{linking}
\eeq
for suitable loops $\gamma$ and $\gamma'$.
Moreover $N=c_v$ and, by  collecting
together all the terms in
the sum of \rf{eqI} and \rf{eqJ}, we get:
\beq
\delta a_2(K)=-{1\over2}
[2\hbox{ln}(K_{1,f}, K) + \hbox{ln}(K_f+K_{2,f}, K_1)+
2\hbox{ln}(K_f+K_{2,f}, K_2)
+\hbox{ln}(K_{1,f}, K_1)],
\label{eqsum}
\eeq
where the knots $K_1$ and $K_2$ (respectively
the framings $K_{1,f}$ and
$K_{2,f}$) represent the two components of the link $K_0$ (respectively
$K_{0,f}$).
In force of the bilinearity of the linking number and of
equation $\hbox{ln} (K_f,K)=0$,
we have $\hbox{ln}(K_{1f},K_1)+\hbox{ln}(K_{1f},K_2)+
\hbox{ln}(K_{2f},K_1)+
\hbox{ln}(K_{2f},K_2)=0$. So \rf{eqsum} can be rewritten in its final
form
\[
\delta a_2(K) = \hbox{ln}(K_{2,f}, K_1) = a_1(K_0).
\]
We have thus recovered the well-known fact that the first
coefficient of the Alexander-Conway polynomial for a two-component link
is the linking number of the two components.

\section{Conclusions}

It is now possible to compare the two existing three-dimensional
topological field theories,
namely the $BF$ theory
and the Chern-Simons-Witten theory (CS).

In both case one can define observables related to
knots and links. And in both cases one obtains link-invariants.

The relation between the Chern-Simons vacuum expectation values of
the observables and the Jones (or HOMFLY) polynomials
can be seen  both as a result of Conformal
Field Theory (as in the Witten original
argument) and (more heuristically) as a consequence
of integration by parts techniques
(as in \cite{CGMM}. See also the more recent \cite{Bru}).

The relation between Alexander-Conway polynomials and the $BF$-theory
(which is  the main result of this paper) relays, for the time being, mainly on
the integration by part techniques. But the structure of the perturbative
series in the $BF$ case, is considerably simpler than in the
CS case. Namely in the $BF$ case we were able to find a
close analytical
expression for the coefficients of the perturbative expansion.

Actual computations look cumbersome, but it
should be possible, with the help of some
computer calculation, to compare directly the $BF$-vacuum expectation values
of some simple knots, with the corresponding
Alexander-Conway polynomial (at least for lower orders).
This has been done
up to the second order of the perturbative expansion, by referring to the
results of \cite{GMM}.

Besides integration by part, other facts
support our claim that the $BF$-theory gives directly the
Alexander-Conway polynomial. The most significant of
these facts is probably the triviality of all the terms of
odd order in the perturbative series. Also the introduction of
connected correlation functions, allows the recovering, in perturbation
theory, of one of the basic property of the Alexander-Conway polynomial,
namely the fact that the polynomial is zero for a link that is
given by the
union of two separated links.

One argument which deserves more work is the comparison between
our multiple integrals (normalized by dividing by the corresponding
integrals for the unknot) and the knot-invariants of the
Kontsevich type \cite{NOI}.

One of the main interests of $BF$ theories
comes from the fact that these theories are topological field
theories that can be defined in both 3 and 4 dimensions.

An observable for 2-knots has already been defined in \cite{CM}.
Techniques completely similar to the one considered in this paper,
can be considered also for the 4-dimensional case. This is
what we are
going to discuss in
a forthcoming paper\cite{NOI2}, where we plan to
connect four-dimensional
topological field theories
to invariants of 2-knots or, maybe, more generally to invariants of
embedded 2-surfaces in 4-manifolds.

\paragraph{Acknowledgements}

We thank J. Baez, J.S. Carter and M. Rinaldi for useful discussions.
P. Cotta-Ramusino
thanks B. Durhuus and J. Dupont for invitations at the Universities
of Copenhagen and Aarhus. M. Martellini thanks J. Ambjorn for an invitation
at the University of Copenhagen.

\newpage
\centerline{\sc Figure Captions}
\begin{description}
\item{\bf Fig.~1} A tangle
\item{\bf Fig.~2} Sum of tangles
\item{\bf Fig.~3} Numerator and denominator of a tangle
\item{\bf Fig.~4} $W_R(({\cal A+B})^D;k)W_R(\bigcirc;k)$
\item{\bf Fig.~5} Decomposition of $\cal A+B$ and $\bigcirc$
\item{\bf Fig.~6} Odd-order graphs (a white
dot denotes the field $A$, a black dot denotes the field $B$)
\item{\bf Fig.~7} An even order V-graph
\item{\bf Fig.~8} Disentangling rule
\item{\bf Fig.~9} Action of the operator ${\cal D}$
\item{\bf Fig.~10} Second order graphs
\item{\bf Fig.~11} Third order graphs
\item{\bf Fig.~12} Fourth order graphs
\item{\bf Fig.~13} Exchange relation for a knot/link
\item{\bf Fig.~14} Exchange relation for a framed knot
(Boldface lines denote the knot, light lines denote the framing)
\end{description}
\end{document}